\documentclass[final,5p,times,twocolumn]{elsarticle}

\begin{document}

	\begin{frontmatter}
		\title{Magnetism,  non-Fermi-liquid behavior and deconfinement in Kondo lattices
		}
		
		
		
		\author{
			V. Yu. Irkhin 		\/\thanks{e-mail: valentin.irkhin@imp.uran.ru}		
		}
		
		\address{M. N. Mikheev Institute of Metal Physics, 620077 Ekaterinburg, Russia}
		\begin{abstract}
			{
				A  formulation of the effective hybridization model is developed which describes  Fermi-liquid Kondo  and  fractionalized Fermi-liquid (deconfinement spin-liquid-like) states with account of possible magnetic ordering. A unified consideration of charge and magnetic fluctuation is performed within perturbation theory in hybridization and within gauge field treatment. The corresponding contributions to specific heat and resistivity are compared.
				
			}
		\end{abstract}
	\end{frontmatter}

\section{Introduction}
Anomalous $f$-systems, which are usually described as Kondo lattices, show unusual behavior of thermodynamic and transport properties, e.g., heavy-fermion behavior \cite{Stewart}: giant  electron specific heat, $\gamma = C/T \sim 1/T_K$ with $T_K$ the Kondo temperature.
Such systems demonstrate typically magnetic instabilities and/or pronounced spin fluctuations \cite{IK97,I17}.
At the same time,  a number of such systems with frustrated magnetic structures demonstrate spin-liquid features with $ \gamma \sim 1/J $ where $J$ is the Heisenberg  exchange interaction, which is small in comparison with electron energies.
One more interesting phenomenon is a possibility of a non-Fermi-liquid behavior where  $\gamma$ diverges at low temperatures.

A scaling theory of the Kondo lattices developed in Ref. \cite{IK97} demonstrates that 
during the scaling procedure the process of magnetic moment compensation
terminates somewhere at the boundary of the strong coupling
region, which can results in the formation of a finite
(although possibly small) saturation moment.
Thus a unified energy scale is established,  both  the effective spin-fluctuation frequency (i.e., $J$) and $T_K$ being strongly renormalized.  Frustration parameters become renormalized too \cite{I19a}.
Therefore. the relative role of magnetism and on-site Kondo effect is determined not only by bare parameters, but also by their mutual renormalizations, 
and a description of the ground state is a complicated problem which  cannot be solved within perturbation theory.

A ground state theory of magnetic Kondo lattices was developed in Refs. \cite{IK91a,IK91,I14} using the mean-field approximation in the pseudofermion representation \cite{Coleman}. 
In the present paper we generalize the approach of Ref. \cite{IK91} and formulate the effective hybridization model describing the competition of magnetic and non-magnetic Kondo and spin-liquid states. We also go beyond the mean-field theory by  taking into account  fluctuations contribution, including gauge field ones.

\section{The model of effective hybridization }

We start from the $s-d(f)$ exchange (``Kondo-Heisenberg'') model.
The corresponding Hamiltonian  reads
\begin{equation}
	H=\sum_{ij\sigma }t_{ij}c_{i\sigma }^{\dagger }c_{j\sigma }-I\sum_{i\alpha \beta }\mathbf{S}_i 
	\mbox {\boldmath $\sigma $}_{\alpha \beta }	c_{i\alpha }^{\dagger }c_{%
		j\beta }+\sum_{ij}J_{ij}\mathbf{S}_{i}%
	\mathbf{S}_{j} \label{H}.
\end{equation}
Here $c_{i\sigma }^{\dagger }$, $c_{j\sigma }$ and $\mathbf{S}_{i}$ are operators for conduction electrons and localized moments, $I$ is the $s-d(f)$ exchange
parameter. 

To construct ground state phase diagrams,  Hartree-Fock and mean-field considerations were developed in Refs.\cite{Kusminskiy,Isaev,Costa}.
The direct Zubarev's decoupling of the $s-d(f)$ exchange interaction was used in these works  to derive the hybridization term  
\begin{eqnarray}
	-I\sum_{\sigma \sigma ^{\prime }}c_{i\sigma }^{\dagger }c_{i\sigma
		^{\prime }}\mbox{\boldmath$\sigma $}_{\sigma \sigma ^{\prime
	}}\mathbf{S}_i 
	\rightarrow V (f_{i\sigma}^{\dagger }c_{i\sigma}+h.c. ) \nonumber \\
	-\frac 12V'\sum_{\sigma \sigma^{\prime }}(f_{i\sigma
	}^{\dagger }\mbox{\boldmath$\sigma $}_{\sigma \sigma ^{\prime
	}}c_{i\sigma ^{\prime }}+h.c.)
		-\frac 4{3I} V^2 +\frac 4{I} V'^2 
	\label{eq:O}
\end{eqnarray}%
This procedure  leads to incorrect value of \textit{exponent} in the Kondo temperature $T_K$  in the paramagnetic state, which is a crucial scale describing heavy-fermion behavior: $T_K=D\exp (2/3I\rho )$ instead of $T_K=D\exp (1/2I\rho ) $ ($D$ is of order of bandwidth, $\rho$ is the density of states at the Fermi level). This violation was demonstrated already in old papers \cite{Takano} (see  discussion in Ref. \cite{Kondo}). 

Therefore we prefer to use the formulation by Coleman and Andrei \cite{Coleman} which was first applied to  the magnetic Kondo lattices in \cite{IK91}.
Making the saddle-point approximation for the path integral describing the
spin-fermion interacting system \cite{Coleman} one reduces the  $s-f$ exchange interaction to the effective hybridization:
\begin{eqnarray}
	-I\sum_{\sigma \sigma ^{\prime }}c_{i\sigma }^{\dagger }c_{i\sigma
		^{\prime }}(\mbox{\boldmath$\sigma $}_{\sigma \sigma ^{\prime
	}}\mathbf{S}_i-\frac 12\delta _{\sigma \sigma ^{\prime }}) \nonumber \\
	\rightarrow f_i^{\dagger }V_ic_i+c_i^{\dagger }V_i^{\dagger
	}f_i-\frac 1{2I}{\rm Sp}{} (V_iV_i^{\dagger }),
	\label{eq:O.2}
\end{eqnarray}%
where the vector notations are used
\[
f_i^{\dagger }=(f_{i\uparrow }^{\dagger },f_{i\downarrow
}^{\dagger }),\qquad c_i^{\dagger }=(c_{i\uparrow }^{\dagger
},c_{i\downarrow }^{\dagger }),
\]
$V$ is the effective hybridization matrix which is determined from a minimum
of the free energy. 
Then we have from the minimization condition the correct value of the Kondo temperature in the paramagnetic state,  
\begin{equation}
	V=2I\sum_{\mathbf{k}}\langle f_{\mathbf{k}\sigma }^{\dagger }c_{%
		\mathbf{k}\sigma }\rangle   \label{eq:O.7}
\end{equation}
and the dispersion of the $f$-states $V^2/t$ at the scale $V^2/D \sim T_K $.

The mean-field Hamiltonian takes the form
\begin{eqnarray}
	H&-&\mu \hat{n} =\sum_{\mathbf{k}\sigma }[(t_{\mathbf{k}}-\mu )c_{\mathbf{k}%
		\sigma }^{\dagger }c_{\mathbf{k}\sigma }+W f_{\mathbf{k}\sigma
	}^{\dagger }f_{\mathbf{k}\sigma}]  \nonumber \\
		&+& \sum_{\mathbf{k}\sigma \sigma'}\sum_{\mu = 0...3}(V^\mu_{\sigma \sigma'}\sigma^\mu_{\sigma \sigma'}c_{\mathbf{k}\sigma }^{\dagger }f_{\mathbf{k}\sigma }+ h.c. )\nonumber \\
		&-&\frac 1{2I}\sum_i {\rm Sp}{} (V_iV_i^{\dagger }) +H_f \label{eq:O.4}
\end{eqnarray}%
where $-W \sim T_K$ is the chemical potential for $f$-states.

Whereas in the usual picture of Kondo singlet formation only scalar 
boson is present (see, e.g., Ref. \cite{Sachdev}), the matrix triplet components create local singlet and triplet states resulting from the Kondo coupling between localized and itinerant spins. Thereby  SU(2) invariance of the $s-f$ exchange interaction is reflected.
Although, as demonstrate numerical calculations \cite{Isaev,Costa}, as a rule the value of $V'$ is probably small in comparison with $V$, this can play important role in magnetization processes (``metamagnetic behavior''), which occur in some heavy-fermion compounds, e.g.,  YbRh$ _{2} $Si$ _{2} $ \cite{Kusminskiy}. 

As for Heisenberg term, we use the pseudofermion representation
\begin{equation}
	\mathbf{S}_i=\frac 12\sum_{\sigma \sigma ^{\prime }}f_{i\sigma
	}^{\dagger }\mbox{\boldmath$\sigma $}_{\sigma \sigma ^{\prime
	}}f_{i\sigma ^{\prime }}
	\label{eq:O.1}
\end{equation}
with the subsidiary condition (constraint)
\begin{equation}
f_{i\uparrow }^{\dagger }f_{i\uparrow }+f_{i\downarrow }^{\dagger
}f_{i\downarrow }=1. \label{constr}
\end{equation}
In the spin-liquid state, $f_{i\sigma}^{\dagger }$ are essentially operators of spinons -- neutral fermions.  After decoupling the Heisenberg term, we obtain their  dispersion on the scale $J$, 
so that the spinon Fermi surface is formed.
The constraint (\ref{constr}) can be enforced by a gauge field. Then in the case of U(1) symmetry we can write down the Lagrangian
\begin{equation}
{\cal L}_f =  \sum_i {f}^\dag_i(\partial_\tau + ia_0)f_i
   -\sum_{ij}\chi_{ij}
\left(e^{ia_{ij'}}f^\dagger_{i\sigma}f_{j\sigma} + \mbox{h.c.} \right)
\end{equation}
where $\chi_{ij} =  J \sum_{\sigma}\langle f^{\dagger}_{i\sigma} f_{j\sigma} \rangle $,
 $a_0$ and $a_{ij}$ are time and coordinate  components of the gauge field. 
Thus the spinons acquire the spectrum $\epsilon_{\bf k} =- \chi_{\bf k}$.
The presence of hybridization results in the formation of a unified ``large'' Fermi surface including both conduction electrons and  $f$-states.




The mean-field treatments \cite{IK91,Sachdev} seem to overestimate the saturated ferromagnetic state, so that  the problem of consistent mean-field description is not solved. An additional effective exchange field was introduced in Ref. \cite{Sachdev} to avoid this difficulty. 
Besides that, the parameter values for phase transitions on the mean-field phase diagrams obtained in Refs.\cite{Kusminskiy,Isaev,Costa} (rather large $I$) were inappropriate for Kondo physics.  
Therefore we pass to a qualitative analysis of fluctuation effects.

To take into account such effects, we can remember that  $f$-states  are in fact described by many-electron Hubbard's projection operators, $f_\sigma \rightarrow X(0,\sigma)=|0\rangle \langle \sigma|$.
Thus we can use slave-particle representations, e.g., $X_i(0,\sigma)=f_{i\sigma} b_i$.
Fluctuations can destroy the condensate picture, so that $\langle b \rangle  = 0$
Then we obtain a spin-liquid-like state where spinon and electron Fermi surfaces are distinct, i.e., a ``small'' Fermi surface. 
 This state was called fractionalized Fermi-liquid (FL$^*$) \cite{Sachdev}. 
 A topological   ``Kondo breakdown'' transition can occur between FL$^*$ and ``usual'' heavy Fermi liquid (FL) states.

To include transverse spin fluctuations, we can use the relation (\ref{constr}) and write down 
$$f_{i\sigma} \rightarrow (f_{i\uparrow }^{\dagger }f_{i\uparrow }+f_{i\downarrow }^{\dagger
}f_{i\downarrow })f_{i\sigma}=-S_i^\sigma f_{i-\sigma}$$ 
with $S_i^\sigma = f_{i\sigma}^{\dagger }f_{i-\sigma ^{\prime }}$ the spin operators
(Instead, we can proceed with equations of motion sequence to obtain similar results.)
This is analogous to using multiplication rules for Hubbard's operators
$	X(0,\sigma)=X(0,-\sigma)X(-\sigma,\sigma)=X(0,-\sigma)S^{-\sigma}.$

Another way to treat all kinds of fluctuations is using the KotliarRuckenstein boson representation \cite{Kotliar86} in  rotationally invariant version \cite{Fresard:1992a}, which was adopted in Ref.\cite{I19} to magnetic state as
\begin{eqnarray}
	f_{i\sigma } &\rightarrow& \frac{1}{\sqrt{2}}\sum_{\sigma
		^{\prime }\mu}f_{i\sigma ^{\prime }} p_{i\mu} \sigma^\mu_{\sigma ^{\prime }\sigma }
\nonumber \\
	&=&\frac{1}{\sqrt{2}}\sum_{\sigma
		^{\prime }}f_{i\sigma ^{\prime }}[\delta _{\sigma \sigma ^{\prime }}p_{i0}+(%
	\mathbf{p}_{i}\mbox{\boldmath$\sigma $}_{\sigma ^{\prime }\sigma })].
	\label{eq:I.88}
\end{eqnarray}
with $p_{i\mu}$ ($\mu=0...3$) the auxiliary Bose operators. This representation satisfies exactly commutation relations for Hubbard's
operators. One more possibility is using  supersymmetric decoupling of the Kondo interaction,  which introduces both fermion and Schwinger boson  operators \cite{Ramires}, so that
\begin{equation}
	\mathbf{S}_{i}   = \frac{1}{{2}}\sum_{i\sigma \sigma '} \mbox {\boldmath $\sigma $}_{\sigma \sigma' }
	(f_{	i\sigma }^{\dagger }f_{		i\sigma ' }^{}+b_{	i\sigma }^{\dagger }b_{		i{\sigma '} }).
	\label{H11}
\end{equation}



\section{Heavy-fermion and non-Fermi-liquid behavior, and magnetism}

As discussed above, in the quantum critical regime, a topological transformation from large Fermi surface (Kondo lattice state) to small Fermi surface is possible, which can be accompanied by  magnetic-order instability.
First we consider the non-magnetic state. 

In the spin-liquid state, gauge field fluctuations play an important role. In particular, there occurs a spinon contribution to specific heat. This contribution retains in FL$^*$ state and in FL state near the topological transition. After integrating out spinons one obtains the gauge field propagator \cite{Sachdev}
\begin{eqnarray}
	D_{\alpha \beta}({\bf q}, i \omega_n) \equiv
	\langle a_\alpha({\bf q}, i \omega_n)a_\beta(-{\bf q}, -i\omega_n)\rangle \nonumber \\
	= \frac{\delta_{\alpha \beta} - q_\alpha q_\beta/q^2}{\Gamma |\omega_n|/q + \chi_f
		q^2 + \rho_s} \,. \label{dprop2}
\end{eqnarray}
Here $\Gamma$ and $ \chi_f$ are   determined
by the details of the spinon dispersion,  $\omega_n$ is an
imaginary Matsubara frequency; $\rho_s \sim b_0^2$ is the boson ``superfluid''  density which is zero in the FL$^*$ state and provides a cutoff for singular gauge-fluctuation contribution. 
Physically, this contribution is due to Landau damping connected with the spinon Fermi surface.
As demonstrate the calculation \cite{Sachdev} with the use of (\ref{dprop2}), the specific heat coefficient $\gamma = C/T$ diverges logarithmically at $T\rightarrow 0$ in the FL$^*$ phase, which means the non-Fermi-liquid behavior (in the 2D case, $C(T) \sim T^{2/3}$). 
At approaching the transition from the FL side, we have $\gamma \sim \ln(1/b_0)$. 
This contribution retains in the insulating phase (in the absence of conduction electrons).

In the quantum critical regime, the fluctuations of order parameter field $b$ are also important. They yield additional contributions to thermodynamic properties. Unlike gauge field (neutral spinon) ones, they contribute also to electronic transport.
The $b$-field fluctuation contribution to the conduction-electron self-energy  can be written in the form \cite{Paul}
\begin{equation}
\Sigma ({\bf k}, i \omega_n) = T \sum_{\Omega_n, \bf q}
G_f({\bf k+q}, i\omega_n + i\Omega_n) D_b({\bf q}, i\Omega_n)
	\label{eq:6.251}
\end{equation}
 where
$G_f({\bf k}, i \omega_n) = 1/(i \omega_n - \epsilon_{\bf k})$ is
the  spinon Green's function. 
Due to mismatch of electron and spinon Fermi surfaces, the decay of bosons into particle-hole pairs becomes possible above an energy
$ E^* $, which can be small if the distance between the two
Fermi surfaces is small. Above this energy the theory  
belongs to the universality class   of the ferromagnet (formally, wave vector $ {\bf Q}=0 $) with the critical exponent  $ z = 3 $ \cite{Vojta}. 
For $E> E^* $ and a spherical Fermi surface
\begin{equation}
	D_{b}(q, \Omega_n) \approx (\rho V^2)^{-1}/[q^2/(4k_F^2) +
	\pi|\Omega_n|/(2\alpha v_F q)]
\end{equation}
with $\alpha$  the ratio of the spinon to the conduction electron bandwidth.

Then at $T> E^* $ one obtains  a $T\ln T $ contribution to
specific heat. Besides that, the fluctuation scattering results in the $ T\ln T $ dependence of  resistivity $R(T)$ in this regime \cite{Paul}.
For $T< E^* $, the behavior of thermodynamic and transport properties depends on that the 
Fermi surfaces intersect or not \cite{Paul1}. In the first case we have $R(T)\sim T^{3/2}$ and $R(T)\sim T \ln T$   in 3d and 2d cases, respectively, as in an antiferromagnet near the quantum phase transition with $z=2$ (although in fact we are in the charge channel). In the second case,  $R(T)\sim T^{2}$.

In the FL states, this  contribution is cut at the Kondo gap of order of $T_K$, which is related to the boson condensate, $T_K \simeq \pi \rho V^2 \langle b \rangle^2 $ in the vicinity of the critical point.






Now we consider the situation of antiferromagnetic (AFM) ordering. 
As it was  concluded in Refs.\cite{Sachdev,Sachdev1},
the FL$^*$ state with the spinon Fermi surface should be unstable with to a  broken-symmetry AFM (spin-density wave, SDW) state. 
Th exotic AFM order  on the spinon Fermi surface  is called AFM$^*$ or  SDW$^*$ phase \cite{Sachdev,Vojta}.  The presence of an SDW$^*$  condensate does not cause a radical change in the structure of the gauge fluctuation in comparison with the FL*  state:
the spinons remain deconfined and  coupled to a gapless U(1) gauge field.
The gauge field excitations coexist with the gapless Goldstone  magnon mode and with a Fermi surface of conduction electrons. Because of the broken translational symmetry, there is no clear difference between small and large Fermi surfaces now.

With increasing $s-d(f)$ coupling, the deconfined phase with small Fermi surface can pass first into usual itinerant AFM state with a large  Fermi surface volume, and then into FL state. The spinon Fermi surface of the FL$^*$-phase  is expected to 
evolve smoothly into the FL region in some vicinity of the topological transition, 
so that the SDW order can continue to FL in the ground state, as discussed in Ref. \cite{IK97}. 
Thus there is no sharp transition between the FL and  FL$^*$ regions,  and there is instead expected to be a large intermediate quantum-critical region \cite{Sachdev}.


To take into account transverse spin fluctuations, we use the rotationally invariant representations discussed in Sect.2  and write down the corresponding self-energy to second order in hybridization,
we derive
\begin{eqnarray}
	\Sigma _{\mathbf{k}}^{(2)}(E)\sim V^2 \sum_{\mathbf{q}}(u_{\mathbf{q}%
	}-v_{\mathbf{q}})^2  \\
\times \left( \frac{1-n_{\mathbf{k+q}}+N_{\mathbf{q}}}{E-\epsilon_{%
			\mathbf{k+q}}-\omega _{\mathbf{q}}}+\frac{n_{\mathbf{k+q}}+N_{\mathbf{q}}}{%
		E-\epsilon_{\mathbf{k+q}}+\omega _{\mathbf{q}}}\right)  \label{S2}
\end{eqnarray} 
where $n_{\mathbf{k}}$ and $N_{\mathbf{q}}$ are Fermi and Bose functions, 
\begin{equation}
\omega _{\mathbf{q}} =2\overline{S}(J_{\mathbf{Q-q}}-J_{\mathbf{Q}%
})^{1/2}(J_{\mathbf{q}}-J_{\mathbf{Q}})^{1/2}  
\end{equation}
is the magnon frequency. and 
\begin{equation}
	u_{\mathbf{q}}^2 =1+v_{\mathbf{q}}^2=\frac 12[1+\overline{S}(J_{\mathbf{q+Q%
	}}+J_{\mathbf{q}}-2J_{\mathbf{Q}})/\omega _{\mathbf{q}}] 
\end{equation}
are coefficients of the Bogoliubov transformation to the magnon operators in the local coordinate system for the ordering with the vector $\mathbf{Q}$.
They satisfy $u_{\mathbf{q}}\mp v_{\mathbf{q}}\cong u_{\mathbf{q+Q}}\pm v_{%
	\mathbf{q+Q}}\propto \omega _{\mathbf{q}}{}^{\pm 1/2}$at $q\rightarrow 0.$ 
This contribution is  similar to that of the usual perturbation theory in the $ s-d(f) $ model \cite{IK95,PhysB}, but works on the scale of the spinon bandwidth.
It is interesting that in the FL$^*$ state the quantity $ V $ plays the role of the $ s-f $ exchange parameter and mediates the RKKY-type interaction between  $ f $-states.
At the same time, the contribution (\ref{S2})  survives in the FL state where a unified Fermi surface arises and  a correlated $s-f$ band with narrow density-of-states peaks is formed. However, in such a situation spectrum $\epsilon_{\mathbf{k}}$ and the prefactor in (\ref{S2}) change.

In the case of weak antiferromagnetism we have an energy scale $T^*= (\Delta/\alpha v_F) J$ ($\Delta $ is the AFM splitting of the  spectrum), so that for $T^{*}<T<J$  the transitions between AFM subbands become singular.
In the general $d=3$ case we have Im$\Sigma (E)\propto E^2$. For $d=2$ we
derive (cf. Ref. \cite{IK95})
\begin{equation}
 	\mathrm{Im}\Sigma ^{(2)}(E)\sim \lambda _{\mathbf{Q}}E \, \mathrm{sgn}E
\end{equation}
with
\begin{equation}
	\lambda _{\mathbf{q}}\sim  V^2 (J_0-J_{\mathbf{Q}})\sum_{%
		\mathbf{k}}\delta (\epsilon_{\mathbf{k}})\delta (\epsilon_{\mathbf{k+q}})  \label{lam}
\end{equation} 

The corresponding intersubband correction to specific heat is 
\begin{equation}
	\delta C_{inter}(T)\sim \sum_{\mathbf{q\simeq Q,}T\leq \omega _{%
			\mathbf{q}}}\lambda _{\mathbf{q}}/\omega _{\mathbf{q}}^2 . \label{intc}
\end{equation} 
In the 2d (or  ``nesting'' 3d) case the integral is logarithmically
divergent at $\mathbf{q\rightarrow Q}$ and the divergence is
cut at $\max (T,T^{*}),$ si that
\begin{equation}
	\delta C_{inter}(T)\sim \lambda _{\mathbf{Q}}T\ln \frac{%
		\bar \omega }{\max (T,T^{*})}.  \label{cinter}
\end{equation}
Thus for $ T> T^*$  we obtain the $T\ln T$-dependence of specific heat. Since the integral in (\ref{intc}) is determined by the magnon spectrum
only, the result (\ref{cinter}) holds also in the case the frustrated (2d-like) magnon spectrum. 
It should be noted that 2d-like spin fluctuations 
near the quantum-critical point are observed in CePdAl  \cite{Zhao1},
CeCu$_{6-x}$Au$_{x}$ \cite{2D}
and YbRh$_{2}$Si$_{2}$ \cite{YbRh2Si2}. 

For general $d=3$ case we have the quadratic temperature dependence of spin-wave
resistivity, $R(T)\propto T^2$.
For the 2d magnon spectrum or ``nested'' 3d situation one obtains 
\begin{equation}
	R(T)\propto T\ln (1-\exp (-T^{*}/T))\simeq T\ln (T/T^{*})  \label{rtlnt}
\end{equation}

The role of magnetic fluctuations depends again on the relation between $J$ and $T_K$ (again, in frustrated and low-dimensional systems with strong short-range the scale $J$ can be considerably larger than $T_N$).
Note that a gauge vector boson
can be introduced  in the vicinity of deconfined quantum critical points, but it is characterized by large damping \cite{vboson}.
 The behavior of thermodynamic properties in quantum-disordered case, including quasi-linear magnetic contributions to specific heat, can be found in Ref. \cite{Sokol}. 
 
Similar to Ref.\cite{IK91}, we can find singular corrections to sublattice magnetization owing to spin-wave damping. The corrections due to hybridization fluctuations in second order in $V$ turn out to be not only of order of $J/T_K$, as in Ref. \cite{IK91}, but also simply of order of unity.

\section{Conclusions}

We have considered various contribution to thermodynamic and transport properties in the vicinity of the  quantum phase transitions in the Kondo lattices.
The gauge field contributions, being connected with the spinon  Fermi surface, result in specific heat and heat transport in the insulating phase.
This dependence can change for other classes of spin liquids, e.g.., for pyrochlores with electron spectrum including Dirac points one has $C(T) \sim T^2 \ln T$ \cite{pyrochlores}.
Experimental data indeed demonstrate considerable $ T $-linear specific heat in narrow-gap insulating and low-carrier  $f$-systems which are usually treated as Kondo lattices, e.g.,
in SmB$ _{6} $ and Ce$ _{3} $Bi$ _{4} $Pt$ _{3} $ \cite{Ce3Bi4Pt3}, CeNiSn \cite{CeNiSn},
YbPtSb \cite{YbPtSb},
Ce$ _{3} $Bi$ _{4} $Pd$ _{3} $ ($\gamma=360 $ mJ/mol K$^2$) \cite{Ce33Bi44Pd33}
CeSbTe ($ T_{N}= 2.6 $ K  41  mJ/mol K$ ^{2} $, $ T_{K} =$ 1.2 K) \cite{CeSbTe},
Yb$ _{3} $Ir$ _{4} $Ge$ _{13} $ \cite{Yb3Ir4Ge13}.	
Moreover, in some such systems, e.g., in CeNi$_{2-\delta}$As$_{2-x} $P$_{x}$ \cite{CeNi}, the non-Fermi-liquid behavior is observed. 

The Kondo breakdown transition  has a topological nature and is connected with a reconstruction of the Fermi surface. Such a transition is expected to occur for the Kondo lattices (unlike the single-impurity Kondo model where only crossovers take place). It is intimately related to magnetic transition, although the detailed  picture is complicated and is not finally understood.
In this connection, the YbRh$ _{2} $Si$ _{2} $ compound is widely discussed. At very low temperatures this system demonstrates a non-trivial competition of Kondo effect, magnetism and spin-liquid behavior in the quantum criticality. For positive chemical pressure  the  Kondo breakdown is observed within the magnetically ordered phase, whereas negative pressure results in their separation, so that intermediate spin-liquid-type ground state occurs in an extended range \cite{YbRh2Si2a}. Related theoretical problems require further investigations.




The research funding from the Ministry of Science and Higher Education of the Russian Federation (the state assignment, theme ``Quantum'' No. 122021000038-7
is  acknowledged.



\end{document}